%% file: main.tex
\documentclass{article}

\usepackage{microtype}
\usepackage{graphicx}
\usepackage{subfigure}
\usepackage{booktabs} % for professional tables
\usepackage{subfiles}
\graphicspath{{pics/}{../pics/}}

\usepackage{hyperref}

\usepackage[accepted]{icml2019}

\usepackage{color}
\icmltitlerunning{Solving Partial Differential Equations with Neural Networks}

\begin{document}

\twocolumn[
\icmltitle{Solving Partial Differential Equations with Neural Networks}



\icmlsetsymbol{equal}{*}

\begin{icmlauthorlist}
\icmlauthor{Juan B. Pedro}{sensioai}
\icmlauthor{Juan Maro\~nas}{upv}
\icmlauthor{Roberto Paredes}{upv}
\end{icmlauthorlist}

\icmlaffiliation{upv}{PRHLT Research Center, Universitat Polit\`ecnica de Val\`encia}

\icmlaffiliation{sensioai}{Sensio AI Solutions}

\icmlcorrespondingauthor{Juan B. Pedro}{sensioai@gmail.com}

\icmlkeywords{Machine Learning, ICML}

\vskip 0.3in
]




\printAffiliationsAndNotice{} % otherwise use the standard text.

\subfile{sections/abstract}
\subfile{sections/intro}

\subfile{sections/related_work}

\subfile{sections/background}
\subfile{sections/method}
\subfile{sections/results}

\subfile{sections/conclusions}

\bibliography{references}
\bibliographystyle{icml2019}

\end{document}

%% file: sections/abstract.tex
\begin{abstract}
Many scientific and industrial applications require solving Partial Differential Equations (PDEs) to describe the physical phenomena of interest. Some examples can be found in the fields of aerodynamics, astrodynamics, combustion and many others. In some exceptional cases an analytical solution
to the PDEs exists, but in the vast majority of the applications some kind
of numerical approximation has to be computed.

In this work, an alternative approach is proposed using neural networks
(NNs) as the approximation function for the PDEs. Unlike traditional numerical methods, NNs have the property to be able to approximate any
function given enough parameters. Moreover, these solutions are continuous and derivable over the entire domain removing the need for
discretization. Another advantage that NNs as function approximations provide is the ability to include the free-parameters in the process of finding the
solution. As a result, the solution can generalize to a range of situations instead of a particular case, avoiding the need of performing new calculations
every time a parameter is changed dramatically decreasing the optimization time.

We believe that the presented method has the potential to disrupt the physics simulation field enabling real-time physics simulation and geometry optimization without the need of big supercomputers to perform expensive and time consuming simulations.
\end{abstract}

%% file: sections/intro.tex
\newcommand{\rparedes}[1]{\textcolor{green}{#1}}

\section{Introduction}
\label{introduction}

 Many scientific and industrial applications require solving Partial Differential Equations (PDEs) to describe physical phenomena such as sound, heat, diffusion, electrostatics, electrodynamics, fluid dynamics, elasticity, or quantum mechanics. Some examples of real-world applications can be found in the fields of aerodynamics for aircraft design, neuroscience for brain activity simulation, etc. Hence, the ability to solve PDEs fast and accurately is an active field of research and industrial interest, and the main motivation of this study \cite{LEV90}.

 In some exceptional cases, yet useful, an analytic solution to the PDEs exists. Nevertheless, the vast majority of interesting real-world applications require some kind of numerical approximation that has to be computed. The numerical simulation of PDEs has been a topic of interest for the scientific community for a long time. Numerous techniques exist today to solve PDEs. Some examples are finite elements, finite difference, finite volumes and Galerkin methods \cite{LEV02} \cite{TOR09} \cite{PIR11}.
 
 All these techniques are based on the idea of domain discretization: divide the computational domain of interest where the PDEs are to be solved and assume a form for the solution on each of these sub-regions. These trail solutions can be as simple as constant values or more elaborated high-order polynomial reconstruction. The global solution is recovered by simply putting together each individual solution. The process of finding the solution to the PDEs consists then on finding the values for the different parameters that minimizes the approximation error. Depending on the numerical technique used, these solutions will provide different properties. Nevertheless, they all share some features directly derived from the discretization process. Namely, they are discontinuous and with limited derivability. Moreover, their accuracy directly depends on the level of precision of the discretization. Numerical schemes used to approximate the solution are constrained by physical effects concerning the rate of change of information between discretized elements. All these aspects usually result in a large amount of discretized elements, which in turn makes some applications unfeasible. 
 
 In order to overcome the previously mentioned limitations of traditional techniques, in this work we explore the use of neural networks (NNs) as the solution approximation for PDEs.

%% file: sections/related_work.tex
\newcommand{\rparedes}[1]{\textcolor{green}{#1}}

\section{Related Work}
\label{related}

Some data-driven approaches to solve PDEs with NNs exist nowadays. In these cases, a lot of simulations are carried out in order to generate a big amount of data that is then used to train NNs in a supervised manner. This is not the approach that we take, since our objective is to remove expensive simulations out of the picture. Our approach is unsupervised and some works that also follow this idea can be found in the literature.

Chiaramonte and Kiener \cite{CHI} use a NN with a single hidden layer and a trial solution to obtain continuous differentiable solution to PDEs. They use the independent variables and a bias as input parameters, and one single 10 output with the optimal values of the trial solution that satisfies the PDE. They test it in the Laplace equation and a conservation law. The results show relatively small errors when compared to the analytical solutions. The advantage of this method is that the obtained solution is a smooth approximation that can be evaluated and differentiated continuously on the domain. Several areas of improvement include adaptive training set generation to reduce training costs and the study of non-uniform discretizations.

Parisi et al. \cite{DAN03} use an unsupervised approach to train a NN to solve PDEs. They take advantage of the universal approximation capabilities of NNs to postulate them as a solution for a given PDE. A single hidden layer perceptron is used as a generic function. The weights are then found by gradient descent optimisation using the original PDE and a set of sample points as error function, using a genetic algorithm for their initialisation. They compared their solution with a traditional method in an unsteady solid-gas reactor problem which relied on spatial discretization obtaining similar accuracy results but at a fraction of the time since once the NN is trained it can find the solution at any given point instantaneously.

Baymani et al. \cite{BAY15} use a NN to solve the Navier Stokes (NS) equations. An analytical solution formed by two parts (one that satisfies boundary conditions and other for the internal domain) is found via optimisation in a feed-forward network with two hidden layers. Results obtained by this method for a two-dimensional steady problem show good agreement with existing data giving smaller errors compared to traditional numerical methods. Furthermore, the solution generated by the NN can be reused at any time.

In \cite{SIR17} Sirignano and Spiliopoulos develop an algorithm similar to Galerkin methods in order to approximate high-order PDEs. Their method is mesh-less, and the NN is trained to satisfy the differential operators and boundary conditions using stochastic gradient descent at randomly sampled spatial points. A similar work is presented in Han et al. \cite{HAN17} who also studied the use of NNs to approximate high-dimensional PDEs.

During the time of preparation of this paper, NVIDIA presented a real-world application of the method at SC19, the annual supercomputing conference, in Denver \cite{NVD19}, min. 43:40. In the conference they present a NN trained to solve the heat flow in a heat sink. By training the NN using the geometry parameters as inputs, they are able to solve the PDEs in a wide range of configurations. They show the real-time heat flow as the geometry changes as well as a new optimal configuration never found before. However, no written work has been found in the literature describing their solution. 

Here we summarise the advantages that solving PDEs with NNs present when compared to traditional methods are:

\begin{itemize}
\item Continuous and derivable solution over the domain, not piecewise discrete (mesh-less).
\item Computational complexity does not increase with the number of sampling points. 
\item Free parameters can be included in the solution, avoiding repeating simulations at different conditions.
\item Once the NN is trained, it can be reused to obtain results instantly.
\end{itemize}

\noindent Although the main disadvantage of this method is the expensive training, the advantages overcome the limitations since once the training is completed, the NN can be reused over and over. 

Our main contribution is the use of a single NN to provide the solution for the PDE in the entire domain. This means we are not assuming any a priori solution form or custom functions to satisfy boundary conditions (a pattern present in almost every reviewed work). Another aspect is the use of NNs with more than 2 layers, an architecture not explored in previous works.

%% file: sections/background.tex
\newcommand{\rparedes}[1]{\textcolor{green}{#1}}

\newcommand{\jmaronas}[1]{\textcolor{red}{#1}}

\section{Background}
\label{methodology}

PDEs are equations that contain unknown multivariate functions and their partial derivatives. This study is restricted to second-order conservation laws of the form 

\begin{equation} \label{convdifeq}
    \phi_t + \nabla \cdot ( {\bf u} \cdot  \phi ) = \nabla \cdot (\Gamma \nabla \phi) 
\end{equation}

Where $\phi$ is the dependent variable, $\phi_t = \partial \phi / \partial t$ is the derivative of $\phi$ w.r.t. to time, $t$, $\nabla = (\frac{\partial}{\partial x}, \frac{\partial}{\partial y}, \frac{\partial}{\partial z})$ is the nabla operator, ${\bf u} = (u, v, w)$ is the velocity and $\Gamma$ is the diffusivity. The independent variables are space, ${\bf x} = (x, y, z)$, and time, $t$. Equation \ref{convdifeq} is known as the convection-diffusion equation and it describes physical phenomena where particles, energy, or other physical quantities are transferred inside a physical system due to two processes: diffusion and convection. Concerning diffusion,$\nabla \cdot (\Gamma \nabla \phi)$, one can assume that $\phi$ is the concentration of a chemical. When concentration is low somewhere compared to the surrounding areas (e.g. a local minimum of concentration), the substance will diffuse in from the surroundings, so the concentration will increase. Conversely, if concentration is high compared to the surroundings (e.g. a local maximum of concentration), then the substance will diffuse out and the concentration will decrease. The net diffusion is proportional to the Laplacian (or second derivative) of concentration if the diffusivity $\Gamma$ is a constant, $\Gamma \nabla^2 \phi$. On the other hand, concerning convection, $\nabla \cdot ( {\bf u} \cdot  \phi )$, imagine the chemical is being transported through a river and we are measuring the water's concentration each second. Upstream, somebody dumps a bucket of the chemical into the river. A while later, you would see the concentration suddenly rise, then fall, as the zone with increased chemical passes by. 

The problems presented in this study as examples will consist on finding the function $\phi({\bf x}, t)$ that satisfies the PDE for a given geometry and initial and boundary conditions. In a traditional method like finite volume, we first divide the computational domain in small regions where the volume average value of $\phi$ at a particular instant is considered

\begin{equation}
    \phi^n_{i} = \frac{1}{V_i} \int_{V_i} \phi({\bf x}, t^n)
\end{equation}

Where $V_i$ denotes the volume of the $ith$ discretized element in the computational domain. One can then obtain the global solution at any time by gathering the individual solutions for all volumes.

\begin{equation}
    \phi^n = \sum_{V_i} \phi^n_{i}
\end{equation}

This function is piece-wise constant and is not derivable, showing the first limitations of traditional numerical methods. Having derivable PDE solutions is important for many applications, since some interesting results are computed with the derivatives, i.e. heat fluxes, mass transfer, etc, A derivable solution will be more accurate than derivatives approximation using piece-wise functions.

In order to update the solution in time, we discretize the different operators in equation \ref{convdifeq} and use a time integration scheme. In the simplest form, using the first-order Euler algorithm,

\begin{equation}
    \phi^{n+1}_{i} = \phi^{n}_{i} - \frac{\Delta t}{V_i} \sum_{f_i} F_{f} A_{f}
\end{equation}

where $\Delta t$ is the time step, $F_{f}$ is the flux of $\phi$ through the face $f$ of the volume $V_i$ and $A_{f}$ is the area of the face. In order to compute the fluxes at the faces, we require additional numerical schemes to approximate these unknown values. Some popular choices are central difference or upwind methods.

In order to close the problem we need two additional elements: the initial condition and the boundary conditions. First, note that an initial condition is just a boundary condition for the time dimension. This is mentioned because initial and boundary conditions are usually treated separately but in our approach they will be treated equally. The initial condition sets the value for $\phi({\bf x}, t=0)$ and in the previously explained time-marching algorithm serves as the first values to start the computation. Boundary conditions, on the other hand, sets the value for $\phi({\bf x} \in D, t>0)$ where $D$ are all the points that lie in the boundary of the domain. Since we cannot see values outside the domain we have to define a particular set of rules to update these points. A lot of boundary conditions exist, but the ones used in our examples are the following:

\begin{enumerate}
    \item Periodic: By assuming periodic conditions the domain folds itself to connect boundaries. 
    \item Dirichlet: For this type of boundary conditions we will fix $\phi$ at boundaries.
    \item Newmann: For this type of boundary conditions we will fix $\nabla \phi$ at boundaries.
\end{enumerate}

Special boundaries may be required for the treatment of walls, inflows or outflows. Note that an initial condition is a Dirichlet boundary condition in the temporal dimension.

Other traditional methods like finite difference or finite elements slightly differ on the methodology used, but the same idea underlies: discretize the computational domain in small regions where a form of the solution is assumed and then recover the global solution by putting them all together. This results in piece-wise solutions which are not derivable. Also, since we use time-marching algorithms, new computations are required every time that we change the free-parameters, initial or boundary conditions.

%% file: sections/method.tex
\newcommand{\rparedes}[1]{\textcolor{green}{#1}}

\section{Methodology}
\label{methodology}

In this section we present a methodology to find a solution for equation \ref{convdifeq} using a neural network. Our goal is to be able to obtain a trained multi-layer perceptron (MLP) that can output $\phi({\bf x},t)$ when ${\bf x}$ and $t$ are set as inputs that also satisfies equation \ref{convdifeq}. 
 The main idea here is that using the independent variables as NN inputs, a forward pass on the network gives us the value of the dependent variables evaluated at that particular point. Since NNs are derivable, we can compute the derivatives of the dependent variables (outputs) w.r.t. the dependent variables (inputs) in order to compute the different derivatives that appear in the original PDEs. With this derivatives, we build a loss function that matches the PDEs and that is used during the training process. If the loss function reaches a near-zero value, we can assume that our NN is indeed a solution to the PDEs. The training process is unsupervised. These solutions are continuous and derivable over the entire domain. An additional interesting property is that NNs allows us to include the free-parameters of the PDEs as part of the solution. As a result, a solution trained for different values of these parameters can generalize to a range of situations instead of a particular case, avoiding the need of performing new calculations every time a parameter is changed. This property is of particular interest in optimisation studies.

Going into more detail, we define a set of points inside our domain in the same way that we would do in traditional methods. We divide these points into two sets, one for 
training the NN and the other for validation during training. We also distinguish between internal points and boundary points. These last will be treated accordingly to the specified boundary conditions (see figure \ref{fig:domain}).

Then, we define the MLP architecture: a number of layers and a number of hidden units in each layer. The number of inputs will be equal to the number of independent variables on the PDEs in addition to any free-parameter that we wish to include. The number of outputs will correspond to the number of unknowns to be solved.

Once we have our training data and the NN defined, the process to find the solution is defined as follows:

\begin{itemize}
    \item For all points, we compute the outputs of the network, $\phi({\bf x},t)$ and the derivatives w.r.t the inputs: $\phi_t$, $\phi_x$, $\phi_{xx}$, etc.
    \item For internal points, we use a loss function that matches our PDE. This is the function we want to optimize for: $\phi_t + \nabla \cdot ( {\bf u} \cdot  \phi ) - \nabla \cdot (\Gamma \nabla \phi) = 0$
    \item For boundary points, since we fix values, we can build a MSE loss function to satisfy the specified condition.
    \item Update the parameters of the NN for each loss function.
\end{itemize}

As it can be seen, the process of training the NN requires the optimization with respect to many loss functions (as many as  PDEs and different boundary conditions) which can be challenging in complex problems and the main limitation found by the authors so far.

%% file: sections/results.tex
\newcommand{\jmaronas}[1]{\textcolor{red}{#1}}
\section{Results}

In this section we illustrate the methodology presented above with two examples. First, a simple one-dimensional advection equation is solved to understand the main process of solving a PDE with a NN. Then, a more challenging two-dimensional case involving first and second order derivatives is solved to showcase the potential of the method, also introducing free parameters as part of the solution.

\label{results}

\subsection{One-dimensional Advection equation}
\label{advection1d}

\begin{figure}
    \centering
    \includegraphics[scale=0.5]{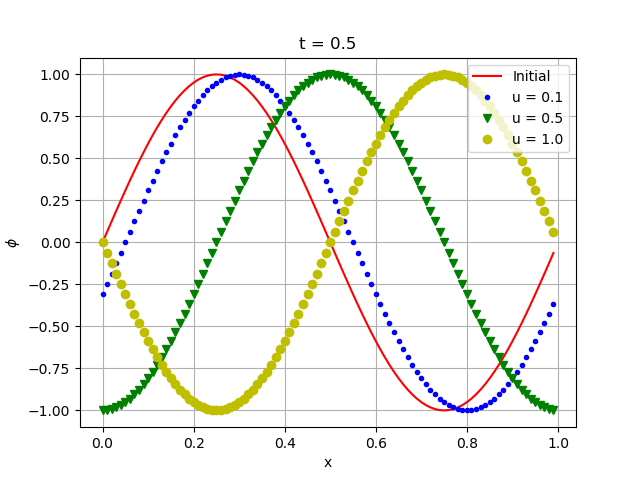}
    \caption{Example of the solution to the one-dimensional advection equation with the initial condition $sin(2\pi x)$}
    \label{fig:sin}
\end{figure}

Consider the one-dimensional advection eqaution which is a simplification of equation \ref{convdifeq} for a one-dimensional inviscid case

\begin{equation}
    \phi_t + u \phi_x = 0
\end{equation}

where $\phi(x,t)$ is the unknown function, $x$ and $t$ are the independent variables, $u$ is a constant parameter and $\phi_t$ and $\phi_x$ are the derivatives of $\phi$ w.r.t $t$ and $x$ respectively. This PDE has analytical solution, which is $\phi(x,t) = \phi(x, x - ut)$. From a physical point of view we can say that the initial condition $\phi(x, t=0)$ is moving in $x$ at the speed $u$. In the case that $\phi(x,t=0) = sin(2\pi x/L)$ the solution is $\phi(x,t)=sin(2\pi (x-ut)/L)$ as illustrated in figure \ref{fig:sin}.

\begin{figure}
    \centering
    \includegraphics[scale=0.2]{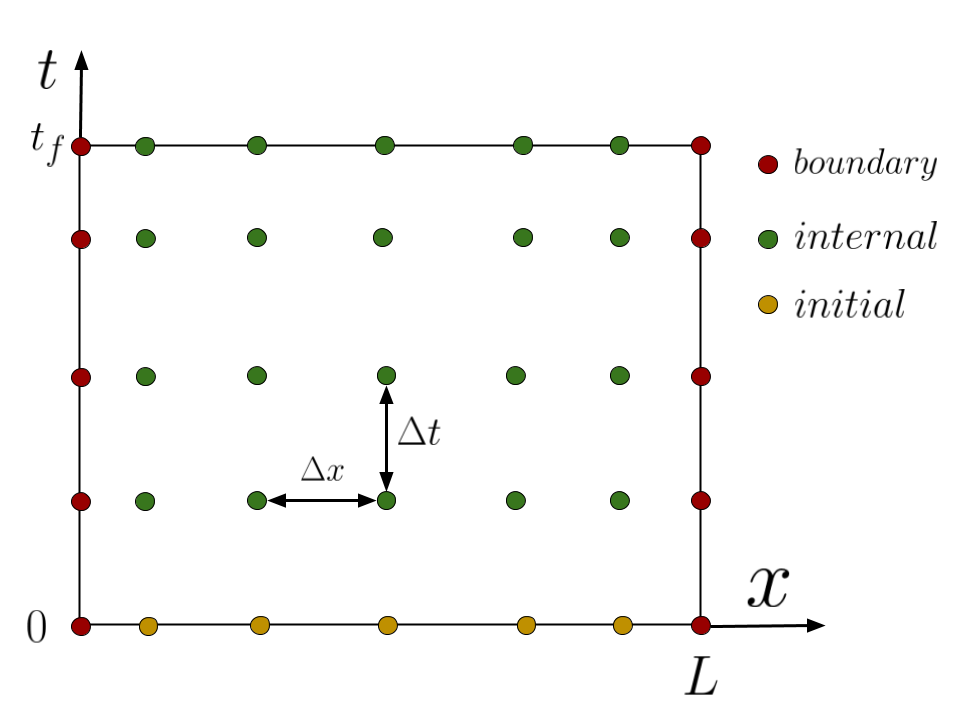}
    \caption{ Example of training points with $N= 5$ and $M= 4$}
    \label{fig:domain}
\end{figure}

To solve the equation we first define a set of points for training. Defining a $\Delta x$ and $\Delta t$ allows us to build a uniform grid of points in the entire domain (see figure \ref{fig:domain}). We define internal and boundary points, which will have different associated loss functions. In this case, the initial condition will use a Mean Square Error loss function that will compare the initial condition (which is known) with the NN output whenever $t=0$. For the spatial boundary condition, we set a periodic condition that will also use a Mean Square Error loss function to compare the solutions at $x=0$ and $x=L$ for any $t$ and force them to be equal.

\begin{figure}
    \centering
    \includegraphics[scale=0.2]{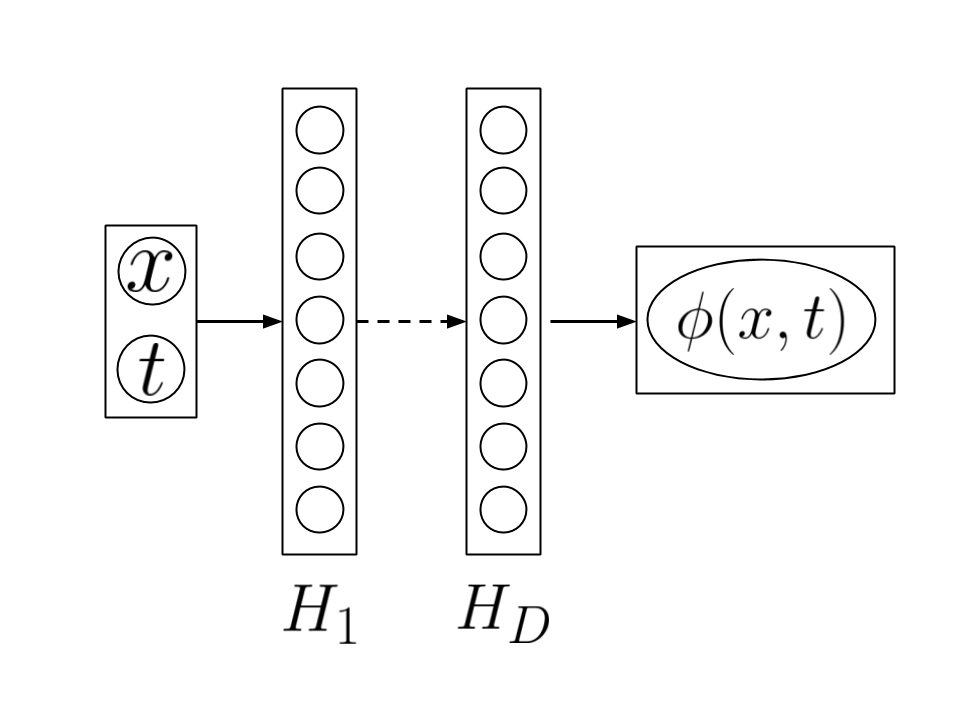}
    \caption{Examlple of solution.}
    \label{fig:nn}
\end{figure}

Also, we define our solution as a multi-layer perceptron with 2 inputs (number of independent variables), $D$ number of hidden layers and 1 output (number of unknowns) as depicted in figure \ref{fig:nn}.

The training goes as follows:

\begin{itemize}
    \item For the internal points, compute the network's outputs: $\phi(0 < x < L,t >  0)$.
    \item Compute the gradients of the outputs w.r.t the inputs: $\phi_x$, $\phi_t$.
    \item Build the loss function for internal points: $L_1 = MSE(\phi_t + u \phi_x$)
    \item Compute outputs for boundary conditions: $\phi(0 < x < L, t=0)$, $\phi(x=0, t)$ and $\phi(x=L, t)$.
    \item Build the loss function for boundary conditions: $L_2 = MSE(\phi(0 < x < L, t=0) - sin(2\phi x/L)$, $L_3 = MSE(\phi(x=0, t) - \phi(x=L, t))$
    \item Update the NN parameters for the different losses.
\end{itemize}

\begin{figure}
    \centering
    \includegraphics[scale=0.4]{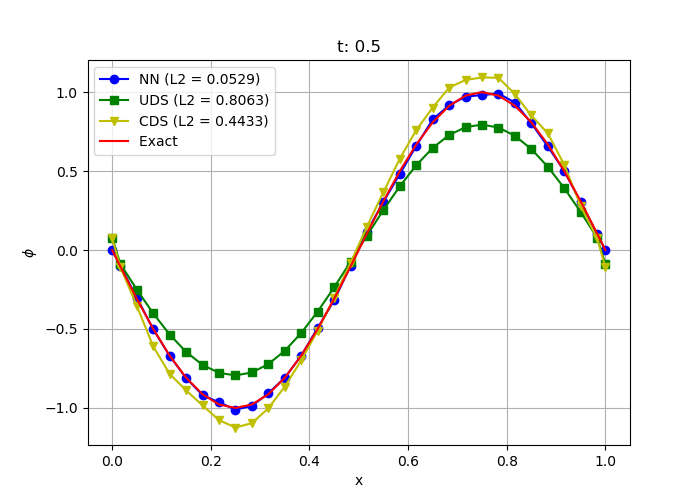} \\
    \includegraphics[scale=0.4]{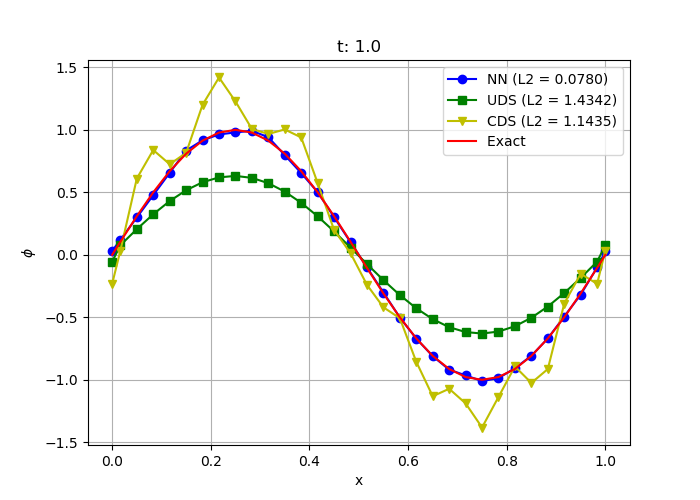}
    \caption{Solution for the one-dimensional advection equation.}
    \label{fig:result_adv1d}
\end{figure}

Results for a 5 hidden layer NN with 32 hidden units and $u = 1$ can be seen in figure \ref{fig:result_adv1d}. Results are compared with traditional FVM with an upwind scheme (UDS) and a central scheme (CDS), both time-integrated with an Euler scheme. Unlike the FVM solution, our trained NN is continuous and derivable over the entire domain. Our experiments show that, for this simple case, we can obtain better results with a much bigger mesh since our methodology is not restricted by physical effects.

\subsection{Two-dimensional Advection-Diffusion equation}
\label{smithHutton2d}

We applied the same methodology introduced in the previous section to solve the viscous Smith-Hutton problem.

\begin{equation}
	(u \phi)_x + (v \phi)_y = \Gamma ( \phi_{xx} + \phi_{yy} )
\end{equation}

\begin{figure}
    \centering
    \includegraphics[scale=0.4]{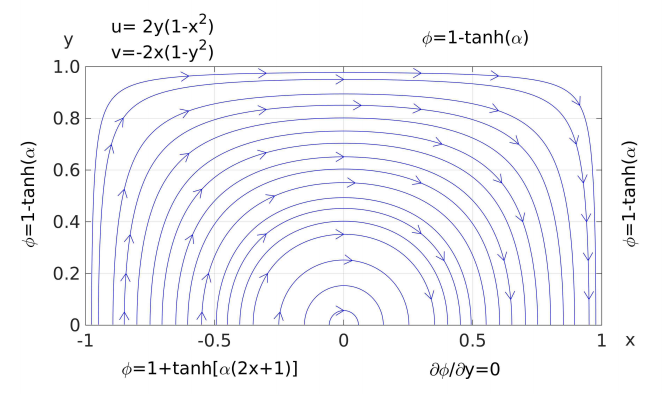} 
    \caption{Smith-Hutton problem domain.}
    \label{fig:sh}
\end{figure}

\begin{figure}
    \centering
    \includegraphics[scale=0.2]{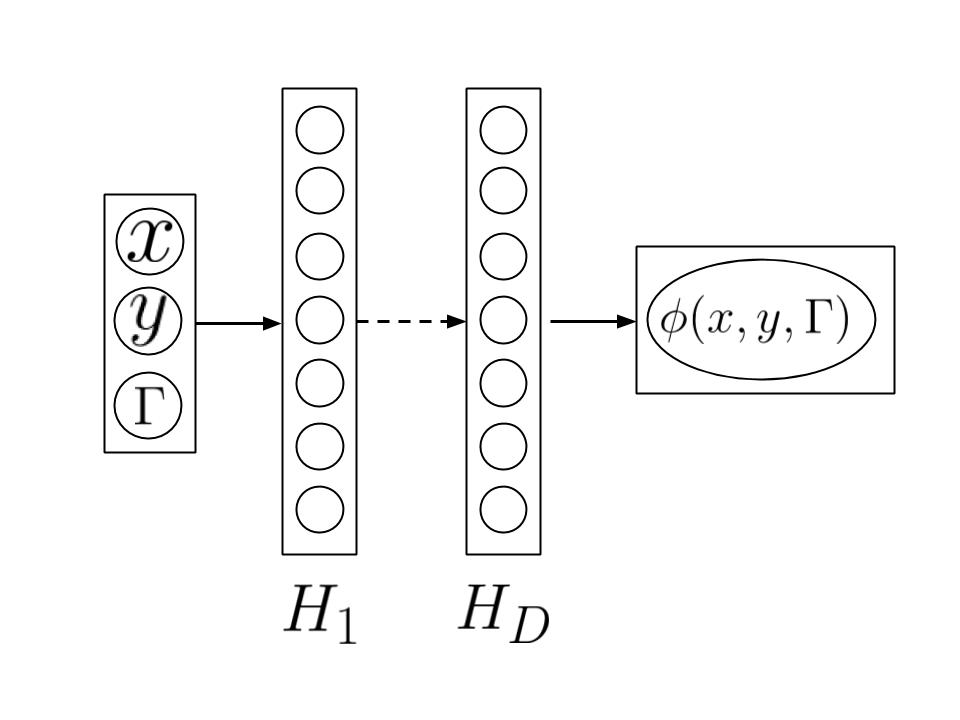} 
    \caption{Smith-Hutton solution architecture.}
    \label{fig:sh_solution}
\end{figure}

In this case we are interested in the steady solution of the two-dimensional advection-diffusion equation in the domain depicted in figure \ref{fig:sh}. The solution used can be seen in figure \ref{fig:sh_solution}.

\begin{figure}
    \centering
    \includegraphics[scale=0.5]{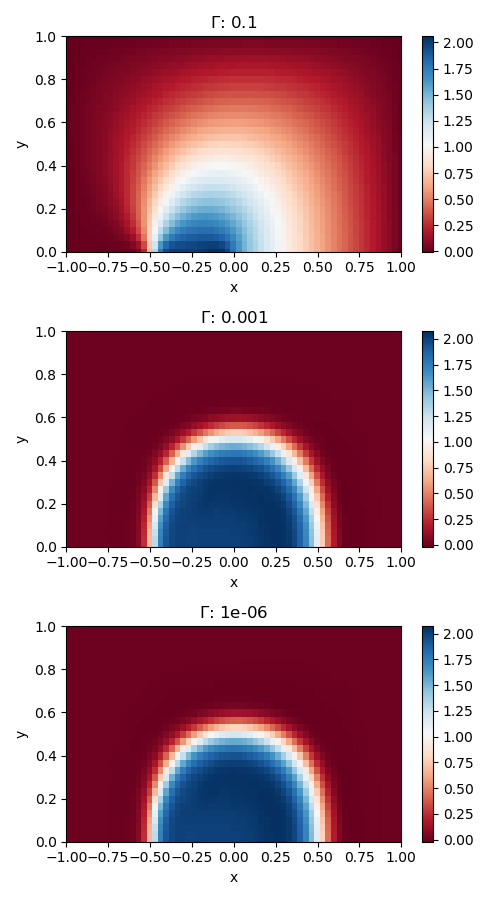} \\
    \includegraphics[scale=0.4]{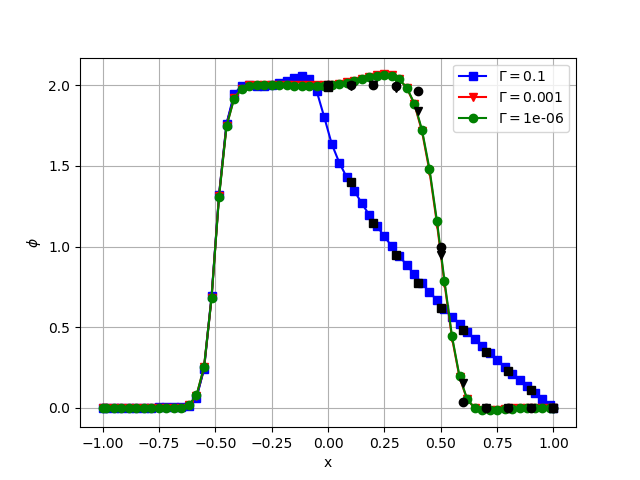}
    \caption{Smith-Hutton solution fields (top) and inlet-outlet profiles (bottom).}
    \label{fig:sh_solution2}
\end{figure}

Results obtained for a $60$x$30$ grid and 3 different values of $\Gamma$ are shown in figure \ref{fig:sh_solution2}. A 4 layer NN with 1024 hidden units in each layer and Sigmoid activation functions was used. Results are in very good agreements with experimental results. It is important to note that all the solutions were obtained by the same NN. Moreover, it is capable of generalize to other conditions not seen during training with an acceptable level of accuracy.

%% file: sections/conclusions.tex
\section{Conclusions}
\label{conclusions}

In this work we have presented a methodology to solve PDEs using NNs. Compared to traditional numerical techniques, our approach is able to provide accurate solutions which are continuous and derivable in the entire domain. Furthermore, free-parameters can be included as NN inputs obtaining solutions in a wide range of conditions. This can dramatically decrease the optimization time of problems where the numerical resolution of PDEs is required.

The proposed methodology consists on using the independent variables of the PDEs as NN inputs. A forward pass on the network gives us the value of the dependent variables evaluated at that particular point. Since NNs are derivable, we can compute the derivatives of the dependent variables (outputs) w.r.t. the dependent variables (inputs) in order to compute the different derivatives that appear in the original PDEs. With this derivatives, we can build a loss function that matches the PDEs and that is used during the training process, which is unsupervised. 

We tested our method in two cases. First, a simple one-dimensional advection equation is solved showing that with a 5 hidden layer NN and 32 hidden units we can provide more accurate solutions than using traditional finite volume methods. Our second case involves the resolution of a two-dimensional advection-diffusion equation. This PDE involves both first and second order derivatives as well as the diffusive parameter introduced as an additional NN input. This results in a NN that is able to provide a continuous solution to the PDE over the entire computational domain and for a wide range of physical conditions. This result dramatically decreases the optimization process and is eye-opening on the multitude of applications that our approach can impact. 

We believe that the presented method has the potential to disrupt the physics simulation field. This powerful technique can be used, for example, to solve the flow over an entire aircraft in a wide range of geometrical and flight conditions. Then it is possible the visualisation of the flow in real time as we change the geometry and physical conditions to obtain the optimal configuration. However, some issues must be addressed before.

When working with systems of PDEs and multiple boundary conditions, our method requires the optimisation of the NN with respect to a lot of loss functions. This results in a big restriction when training the NN. Also, some training strategies can be devised to reach better results, such as mesh refinement or data augmentation tailored to our application.